\documentclass[prl,twocolumn,floatfix,showpacs,epsf,amssymb,longbibliography]{revtex4-1}
\usepackage[utf8]{inputenc}
\usepackage[english]{babel}
\usepackage{amsmath}
\usepackage{amsfonts}
\usepackage{amssymb}
\usepackage{graphicx}
\usepackage{float}
\usepackage{xcolor}
\usepackage{subfigure}
\usepackage{braket}
\usepackage{hyperref}
\hypersetup{pdfborder=0 0 0,colorlinks=true,citecolor=blue,linkcolor=blue}
\usepackage{mciteplus}

\begin{document}

\title{A doping-dependent switch from  one- to two-component superfluidity at \\ high temperatures
in coupled electron-hole van der Waals heterostructures}
\author{Sara Conti$^{1,2}$, Matthias Van der Donck$^{2}$, Andrea Perali$^{3}$, Francois M. Peeters$^{2}$, and David Neilson$^{1,2}$}
\affiliation{
\mbox{$^1$Physics Div., School of Science \& Technology, Universit\`a di Camerino, 62032 Camerino (MC), Italy}\\
\mbox{$^2$Department of Physics, University of Antwerp, Groenenborgerlaan 171, B-2020 Antwerpen, Belgium}\\
\mbox{$^3$Supernano Laboratory, School of Pharmacy, Universit\`a di Camerino, 62032 Camerino (MC), Italy}}
\begin{abstract}
The hunt for high temperature superfluidity has received new impetus from the discovery of atomically thin stable materials.  
Electron-hole superfluidity in coupled MoSe$_2$-WSe$_2$ monolayers is investigated using 
a mean-field multiband model that includes the band splitting  caused by the strong spin-orbit coupling.  
This splitting leads to a large energy misalignment of the electron and hole bands which is strongly modified 
by interchanging the doping of the monolayers.  
The choice of doping determines if the superfluidity is tuneable from one- to two-components.  
The electron-hole pairing is strong, with high transition temperatures in excess of $T_c\sim 100$ K.
\end{abstract}
 
\maketitle
 
Recently strong signature of electron-hole superfluidity was reported in double bilayer graphene (DBG) \cite{Burg2018}, 
in which an $n$-doped bilayer graphene was placed in close proximity with a $p$-doped bilayer graphene, 
separated by a very thin insulating barrier to block recombination.  
The transition temperature is very low, $T_c\sim 1$ K. This can be traced back to the 
very strong interband screening \cite{Conti2019} due to bilayer graphene's tiny band gap \cite{Zhang2009}.

Monolayers of the Transition Metal Dichalcogenides (TMDC) MoS$_2$, MoSe$_2$, WS$_2$, and WSe$_2$ 
are semiconductors with large and direct bandgaps,  $E_g\gtrsim1$ eV \cite{Mak2010,Jiang2012}  that 
make interband processes and screening negligible.  
The effective masses in their low-lying nearly parabolic bands, are larger than in bilayer graphene, resulting also in much stronger coupling of the electron-hole pairs \cite{Fogler2014}.  

Because of the strong spin-orbit coupling, the heterostructure MoSe$_2$-hBN-WSe$_2$, with one TMDC monolayer $n$-doped and the other $p$-doped,  
is an interesting platform for investigating novel multicomponent effects for electron-hole superfluidity \cite{Rivera2015,Ovesen2019, Forg2019}.
The hexagonal Boron Nitride (hBN) insulating layer inhibits electron-hole recombination \cite{Britnell2012a}, and avoids hybridization between the MoSe$_2$ and WSe$_2$ bands.

\begin{figure}[t]
\centering
\includegraphics[trim=1.2cm 1cm 2.0cm 1.0cm, clip=true, width=1\columnwidth]{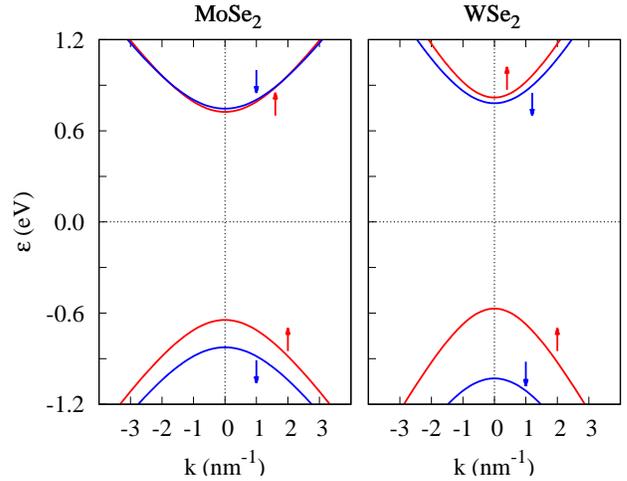} 
\caption{(Color online) The low-lying band structures of monolayer MoSe$_2$ and WSe$_2$ centred in the $K$ valley. 
Red and blue lines are for the opposite spins.  The spin configuration is opposite in the two valleys \cite{Xiao2012}. 
}
\label{fig:TMD_bands}
\end{figure}

Table \ref{table:TMDs} gives the parameters for the MoSe$_2$ and WSe$_2$ monolayers, and Fig. \ref{fig:TMD_bands} shows their low-lying band structures.   
The splitting of the conduction and valence bands by spin-orbit coupling into multibands consisting of two
concentric parabolic spin-polarised subbands, makes superfluidity in double TMDC monolayers 
resemble high-$T_c$ multiband superconductivity.  
Multiband superconductivity is emerging as a complex quantum coherent phenomenon with 
physical outcomes radically different, or even absent, from its single-band counterparts \cite{Bianconi2013}. 
There are close relations with multiband superfluidity in ultracold Fermi gases \cite{Shanenko2012} 
and with electric-field induced superconductivity at oxide surfaces \cite{Mizohata2013,Singh2019}.
\setlength\belowcaptionskip{-0.3cm}
\begin{table}[t]
\centering 
\begin{tabular}{|c | c | c | c | c | c|} 
\hline  
TMDC      & a (nm) & t (eV) & $E_g$ (eV) & $\lambda_c$ (eV) & $\lambda_v$ (eV) \\ [0.5ex] 
\hline 
\hline
MoSe$_2$ & 0.33	 & 0.94   & 1.47       &  -0.021          &  0.18 \\ [0.5ex] 
\hline 
WSe$_2$  & 0.33	 & 1.19   & 1.60       &  0.038           &  0.46 \\ [0.5ex] 
\hline 
\end{tabular}
\caption{TMDC monolayer lattice constant (a), hopping parameter (t), band gap ($E_g$),  and splitting of conduction band ($\lambda_c$) and valence band ($\lambda_v$)  
by spin-orbit coupling \cite{Xiao2012,Zhu2011,Kosmider2013}. } 
\label{table:TMDs} 
\end{table}

\begin{figure*}[t]
\centering
\includegraphics[trim=1.0cm 0.0cm 1.0cm 0.3cm, clip=true, width=1\textwidth]{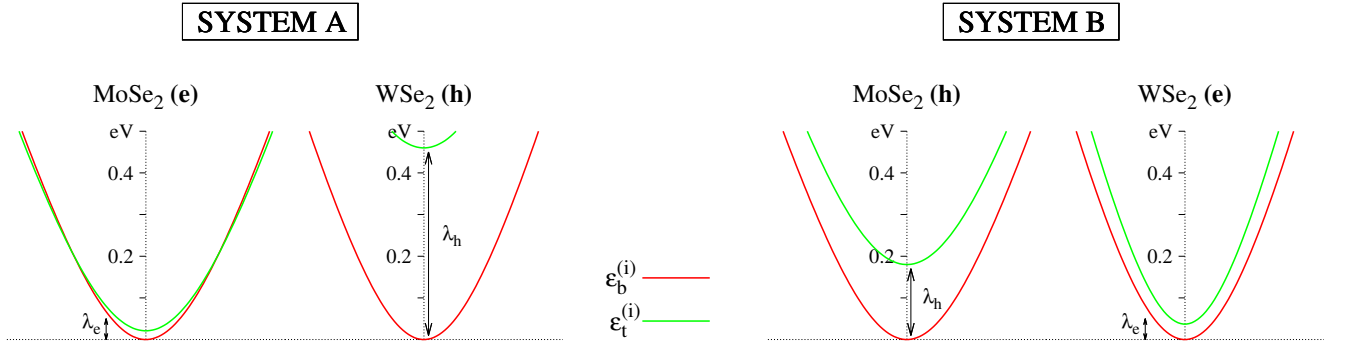}
\caption{(Color online) Subbands of systems A and B (see text) centred in the $K$ valley.  
For the $p$-doped monolayer, the valence band has been mapped into a conduction band using the standard particle-hole transformation.  
The bottom electron $\varepsilon^{e}_b(k)$ and hole $\varepsilon^{h}_b(k)$ subbands have been aligned.  
Zero energy is set at  $\varepsilon^{e}_b(k=0)$.}
\label{fig:systems}
\end{figure*}

Table \ref{table:TMDs} shows that the spin splitting of the valence bands $\lambda_v$ is an order of magnitude larger 
than the spin splitting of the conduction bands $\lambda_c$.  
This results in a misalignment between the electron and hole bands, as shown in Fig.\ \ref{fig:systems}. 
(For the $p$-doped monolayer, we are using the standard particle-hole mapping of the valence band to a conduction band, 
with positively charged holes filling conduction band states up to the Fermi level. Thanks to the large band gaps, we only need to consider conduction band processes \cite{Conti2017,Conti2019}.)
A Coulomb pairing interaction, in contrast with conventional BCS pairing, has no dependence on the electron and hole spins. Therefore for each monolayer, we label the bottom and top conduction subbands by $\beta=b$ and $\beta=t$.
Due to the large valley separation in momentum space, intervalley scattering is negligible, so the effect of the two valleys appears only in a valley degeneracy factor, $g_v=2$. 

We will find the misalignment strongly affects the electron-hole pairing processes, and that
due to the very different misalignments of the bands (Fig.\ \ref{fig:systems}), 
the $n$-doped-MoSe$_2$ with $p$-doped-WSe$_2$ (denoted as system A)
has markedly different properties
from the $p$-doped-MoSe$_2$ with $n$-doped-WSe$_2$ (system B). 

The multiband electron-hole Hamiltonian is, 
\begin{equation}
\begin{aligned}
&H=\sum_{k,\beta} \;\left\{\xi^{(e)}_\beta(k)  \,c^{\dagger}_{\beta,k} \, c_{\beta,k} +
\xi^{(h)}_\beta(k) \,d^{\dagger}_{\beta,k} \, d_{\beta,k} \right\} \\ 
 &+ \!\!\!\sum_{\substack{k,k',q\\ \beta,\beta'}} \!\!\! V^D_{k\, k'}
\,c^{\dagger}_{\beta,k+ q/2}
\,d^{\dagger}_{\beta,-k+ q/2}
\,c_{\beta',k'+ q/2}
\,d_{\beta',-k'+ q/2} 
\end{aligned}
\label{eq:Hamiltonian}
\end{equation}
For the $n$-doped monolayer, $c^{\dagger}_{\beta,k}$  and $c_{\beta,k}$ are the creation and annihilation operators for electrons in conduction subband $\beta$,
while for the $p$-doped monolayer, $d^{\dagger}_{\beta,k}$ and $d_{\beta,k}$ are the corresponding operators for holes.
The kinetic energy terms are $\xi^{(i)}_\beta(k) = \varepsilon^{(i)}_\beta(k)-\mu^{(i)}$ 
where $\varepsilon^{(i)}_\beta(k)$ is the energy dispersion for the $i=e,h$ monolayer \cite{Vanderdonck2018}.  
Because of the small difference between electron and hole effective masses, we assume bands of the same curvature,
and so since we consider only equal carrier densities $n^e=n^{h}=n$, 
for the chemical potentials $\mu^{(e)}=\mu^{(h)}\equiv \mu$. 
$V^D_{k\, k'}$ is the bare attractive Coulomb interaction between electrons and holes in opposite monolayers 
separated by a barrier of thickness $d$, 
\begin{equation}
V^D_{k\, k'} = -V^S_{k\, k'} e^{-d|\textbf{k}-\textbf{k}'|}\ , \quad V^S_{k\, k'} = \frac{2\pi e^2}{\epsilon}\frac{1}{|\textbf{k}-\textbf{k}'|}\ ,
\label{eq:bare_interacions}
\end{equation}
where $V^S_{k\, k'}$ is the bare repulsive Coulomb interaction between carriers in the same monolayer.
 
In principle there are four possible electron-hole pairings, corresponding to four superfluid 
condensates \cite{Shanenko2015} $\{\beta\beta'\}$. The first index $\beta$ refers to the electron subbands 
and the second $\beta'$ to the hole subbands. 
We find that the $\{bt\}$ and $\{tb\}$ cross-pairing make negligible contributions to the condensates, 
so for simplicity, we confine our attention to the mean-field equations for the superfluid gaps $\Delta_{bb}(k)$ and $\Delta_{tt}(k)$. 
Since there are no spin-flip scattering processes, Josephson-like pair transfer is forbidden.
At zero temperature these gap equations are (see Appendix),
\begin{align}
\Delta_{bb}(k) &=-\frac{1}{L^2}\sum_{k'} F^{bb}_{kk'} \, V^{eh}_{k\, k'} \,\frac{\Delta_{bb}(k')}{2 E_b(k')} \quad , \vspace*{2mm} \label{eq:gapbb} \\
\Delta_{tt}(k) &=-\frac{1}{L^2}\sum_{k'} F^{tt}_{kk'} \, V^{eh}_{k\, k'} \,\frac{\Delta_{tt}(k')}{2 E_t(k')}  \theta[E^-_t(k')]
\ .
\label{eq:gaptt}
\end{align}
$E_\beta(k)=\sqrt{\xi_\beta(k)^2 + \Delta^2_{\beta\beta}(k)}$ is the quasi-particle excitation energy for subband $\beta$, with $\xi_\beta(k)=(\xi^{(e)}_\beta+\xi^{(h)}_\beta)/2$. 
$E^\pm_t(k)=E_t(k)\pm\delta \lambda$ with \textbf{$\delta\lambda= (\lambda_h-\lambda_e)/2$}. 
$\lambda_h$ is the spin-splitting of the conduction band of the $p$-doped monolayer, 
and $\lambda_e$ the corresponding spin-splitting for the $n$-doped monolayer, with values taken from Table \ref{table:TMDs}. 
$\theta[E^-_t(k)]=1-\mathit{f}[E^-_t(k),0]$ is a step function associated with the zero temperature Fermi-Dirac distribution.
$F^{\beta\beta}_{kk'}=|\Braket{\beta k|\beta k'}|^2$ is the form factor that accounts for the overlap of single-particle states in $k$ and $k'$ 
for subbands $\beta$ in opposite monolayers \cite{Lozovik2009} (see Appendix). 

$V^{eh}_{kk’}$ in Eqs.\ (\ref{eq:gapbb}-\ref{eq:gaptt}) is the screened electron-hole interaction. 
We use the linear-response random phase approximation for static screening in the superfluid state \cite{Conti2019}, 
\vspace{-0.5cm}
\begin{widetext}
\begin{equation}
V^{eh}_{k\, k'}  = 
\frac{V^D_{k\, k'}  + \Pi_a(q)[(V^S_{k\, k'} )^2-(V^D_{k\, k'} )^2]}
{1- 2[V^S_{k\, k'}  \Pi_n(q) + V^D_{k\, k'}  \Pi_a(q)] + [\Pi_n^2(q) - \Pi_a^2(q)][(V^S_{k\, k'} )^2 - (V^D_{k\, k'} )^2]} \ , 
\label{eq:VeffSF}
\end{equation}
\end{widetext}
where $q=|\textbf{k}-\textbf{k}'|$. $\Pi_n(q)$ is the normal polarizability in the superfluid state and
$\Pi_a(q)$ is the anomalous polarizability \cite{Lozovik2012,Perali2013}, which is only non-zero in the superfluid state.
$\Pi_n(q)$ depends on the population of free carriers (see Appendix).   
$\Pi_a(q)$, with opposite sign, depends on the population of electron-hole pairs.  
The combined effect of $\Pi_n(q)$ and $\Pi_a(q)$ is that a large superfluid condensate fraction of strong-coupled and approximately neutral pairs is associated with very weak screening \cite{Neilson2014}. This is because of the small remaining population of charged free carriers available for screening.

Equation (\ref{eq:gapbb}) has the same form as for a decoupled one-band system,because the two $b$ bands are aligned \cite{Kochorbe1993}.
In contrast, Eq.\ (\ref{eq:gaptt}) shows explicitly the effect of misalignment of the $t$ bands (Fig. \ref{fig:systems}) through the term $\theta[E^-_t(k')]\equiv\theta[\sqrt{\xi_t(k)^2 + \Delta^2_{tt}(k)}-\delta \lambda]$.
This can only drop below unity at higher densities where the pair coupling strength is weak compared with the misalignment. 

For a given chemical potential $\mu$, the carrier density $n$ of one monolayer is determined as a sum of the subband carrier densities $n_{b}$ and $n_{t}$ by,
\begin{align}
n&= g_s g_v \sum_{\beta=b,t} n_{\beta} \label{eq:density}\\
n_{b}&=\frac{1}{L^2}\sum_k v^2_{b}(k)\label{eq:densityb} \\
n_{t}&=\frac{1}{L^2}\sum_k v^2_{t}(k)\theta[E^+_t(k)] + u^2_{t}(k)(1-\theta[E^-_t(k)])
\label{eq:densityt}
\end{align}
where $v^2_{\beta}$ and $u^2_{\beta}$ are the Bogoliubov amplitudes for the subbands $\beta$ (see Appendix). Because of the spin polarisation in the valleys, the spin degeneracy is $g_s=1$.

The regimes of the superfluid crossover are characterized by the superfluid condensate fraction $C$ \cite{Salasnich2005, LopezRios2018}.  
$C$ is defined as the fraction of carriers bound in pairs relative to the total number of carriers.  
For $C>0.8$ the condensate is in the strong-coupled BEC regime, for $0.2\leq C \leq 0.8$ in the crossover regime, and for $C<0.2$ in the BCS regime.  
In our system, the two condensate fractions are given by,
\begin{equation}
C_{\beta\beta}=\frac{\sum_{k} u_{\beta}^2(k)\; v_{\beta}^2(k)}{\sum_{k} v_{\beta}^2(k)}.
\label{eq:CF}
\end{equation}

\begin{figure}[h!]
\centering
\includegraphics[width=1\columnwidth]{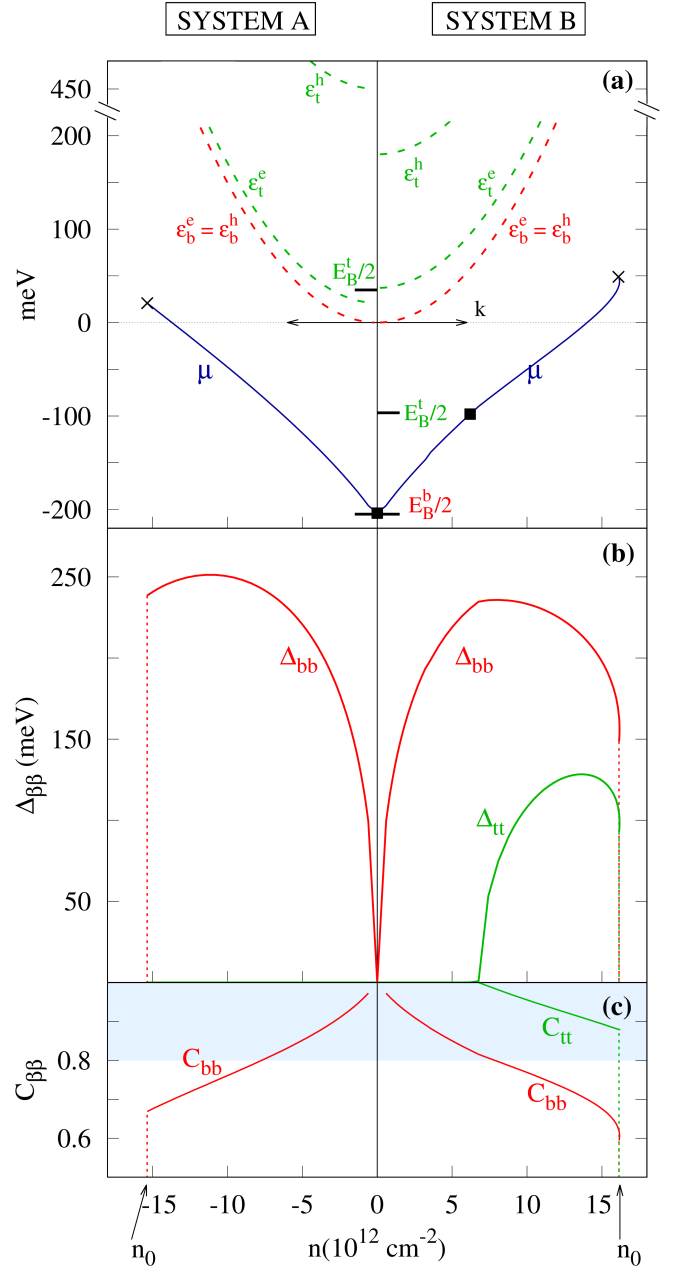} 
\caption{(Color online) (a) Chemical potential as function of density $n$ of WSe$_2$. 
Positive density corresponds to system A, negative density to system B. 
For reference, the energy bands are shown as a function of $k$ with the same energy scale. 
The bound state energies $E^b_B/2$, $E^t_B/2$ are also indicated with respect to the bands. 
(b) The maximum of the superfluid gaps $\Delta_{bb}$ $\Delta_{tt}$ as a function of $n$.
(c) Corresponding condensate fraction $C_{bb}$ and $C_{tt}$. The blue shaded area is the BEC regime. }
\label{fig:result}
\end{figure}

Figure \ref{fig:result}(b) shows the dependence on WSe$_2$ electron density of the maximum of the superfluid gaps $\Delta_{\beta\beta}=\max_k\Delta_{\beta\beta}(k)$ 
for the b and t bands (Eqs.\ (\ref{eq:gapbb}-\ref{eq:gaptt})) in systems A and B. 
We took equal effective masses $m^*_e=m^*_h=0.44 m_e$, a barrier thickness $ d=1$ nm, and dielectric constant $\epsilon=2$,  
for monolayers encapsulated in few layers of hBN \cite{Kumar2016}.

Figure \ref{fig:result}(c) shows the evolution of the condensate fractions (Eq.\ (\ref{eq:CF})) as a function of density, and Fig.\ \ref{fig:result}(a) the evolution of the chemical potential.

We see in Fig.\ \ref{fig:result}(b) that the form of $\Delta_{bb}$ is similar for systems A and B.  
At low densities the system is in the strong coupled BEC regime, with condensate fraction $C_{bb} >0.8$. 
At these densities the $\{bb\} $ pairing is to a deep bound state with binding energy $E_B^{b}\sim 400$ meV below the bottom of the $b$ band \cite{Randeria1990,Pistolesi1994} .  
The chemical potential is $\mu\sim -E_B^{b}/2\,$ (Fig.\ \ref{fig:result}(a)).
With increasing density, $\Delta_{bb}$ increases and then passes through a maximum. $\mu$ also increases and approaches zero.  
Eventually, $\Delta_{bb}$ drops sharply to zero at a superfluid threshold density $n_0$. 
For $n>n_0$, the screening of the pairing interaction is so strong that it kills superfluidity \cite{Perali2013}.

In contrast, $\Delta_{tt}$ is only non-zero in system B. 
At low density, $\Delta_{tt}=0$ also in system B, since the pairing population is zero. This is because the chemical potential $\mu$ at these densities 
lies below the isolated bound state associated with the $t$ bands, located at energy $E_B^{t}=E_B^{b}-(\lambda_e+ \lambda_h)$.
It is only when $\mu$ passes above $-E_B^{t}/2$ that this state can be populated, so $\Delta_{tt}$ can  become non-zero.  
Further increasing the density increases the $\{tt\}$ pair population, $\Delta_{tt}$ increases and then passes through a maximum. 
When $\mu$ becomes positive, the build up of free carriers, as evidenced by $C_{bb} < 0.8$ in Fig.\ \ref{fig:result}(c), 
combined with the misalignment of the $t$ bands, starts to significantly weaken the effective electron-hole screened interaction. 
Eventually screening kills the superfluidity in both $\{bb\}$ and $\{tt\}$ channels at the same threshold density.

We see in Fig.\ \ref{fig:result}(b) that the behavior of $\Delta_{tt}$ in systems A and B is completely different. 
In system A the chemical potential remains below the isolated bound state $E_B^{t}$ associated with the $t$ bands 
over the full range of densities up to $n_0$. With $\mu$ lying below $E_B^{t}$, the population of pairs in the $\{tt\}$ channel remains zero. 
The only difference between system A and B is the choice of doping which results in the markedly different misalignment of the $t$ bands, leading to one-component or two-components superfluidity. 

In Fig.\ \ref{fig:result}(c), we note that the threshold densities $n_0$ for the superfluidity are much larger than the threshold densities 
$n_0\sim 8\times 10^{11}$ cm$^{-2}$ in double bilayer graphene \cite{Burg2018,Conti2019}, 
and the $n_0\sim 4\times 10^{12}$ cm$^{-2}$  predicted for double layer phosphorene \cite{Saberi2018}.  
$n_0$ is large for the double TMDC monolayers for two main reasons: (i) the large effective masses of the electrons 
and holes means a large effective Rydberg energy scale, thus large superfluid gaps $\Delta$ that strongly suppress 
the screening; (ii) the large TMDC monolayer bandgaps $E_g$ eliminate valence band screening, 
making the electron-hole pairing interaction very strong \cite{Conti2019}.  

These large threshold densities in the double TMDC monolayers lead to high Berezinskii-Kosterlitz-Thouless transition temperatures $T_{KT}$ \cite{Kosterlitz1973}.  The monolayers have near parabolic bands, 
so we can approximate \cite{Benfatto2008,Botelho2006}, 
\begin{equation}
T_{KT} = \frac{\pi}{2} \rho_s(T_{KT}) \simeq  n\, \frac{\pi\hbar^2}{8 g_s g_v m^*}  \ .
\label{eq:T_KT1}
\end{equation}
$\rho_s(T)$ is the superfluid stiffness.
Equation (\ref{eq:T_KT1}) gives transition temperatures for systems A and B 
at their threshold densities of $T_{KT}^A= 110$ K and $T_{KT}^B= 120$ K.

The strikingly different behavior of $\Delta_{tt}$ in the two systems is a new and remarkable effect that can be probed using angle-resolved photoemission spectroscopy (ARPES) \cite{Rist2013}.
ARPES measures the spectral function, which in a one-component superfluid state like system A will have a single  peak centred at a negative frequency corresponding to $\Delta_{bb}$.
However in system B, when it switches from  one-component  to two-components superfluidity, 
two peaks associated with the gaps $\Delta_{bb}$ and 
$\Delta_{tt}$  will appear in the spectral function 
at negative frequencies \cite{Miao2012}.  
Other experimental techniques that can be used to detect the presence or absence of the second gap  $\Delta_{tt}$ are Andreev reflection spectroscopy \cite{Daghero2014,Kuzmicheva2016} and scanning tunneling microscopy (STM) \cite{Yin2015}.

The large gaps at zero temperature and in the BCS-BEC crossover regime should lead to pseudogaps in the single-particle excitation spectra \cite{Perali2002} above $T_{KT}$, that persist up to temperatures of the order of the zero temperature gaps. These could also be detected by theARPES and STM. System B at densities where both the superfluid components are close to their maximum gaps would favour large pseudogaps, while configurations with one large gap and one small or zero gap would lead to screening of superfluid fluctuations and suppression of the pseudogap \cite{Salasnich2019}.

In summary, we have investigated multicomponent effects for electron-hole multiband superfluidity 
in $n$-$p$ and $p$-$n$ doped  MoSe$_2$-hBN-WSe$_2$ heterostructures (systems A and B, respectively).  
Both systems are multiband and can stabilize electron-hole superfluidity at temperatures above $100$ K. 

Surprisingly we find that only in system B can superfluidity have two components. 
For both systems we would have expected to be able to tune from one- to two-
component superfluidity by increasing the density, as recently observed in multiband superconductors \cite{Singh2019}, and this is indeed the case for system B.  However for 
system A, the very large misalignment of the electron and hole top bands, means that there are no carriers available for 
pairing in the topmost band before screening has become so strong that it completely suppresses superfluidity.  Therefore 
only one-component superfluidity is possible in system A.  This is a remarkable result: activation of the second-component 
of the superfluidity in this heterostructure depends crucially on the choice of which TMDC monolayer is 
$n$-doped and which $p$-doped.

\textit{ After completion of this paper we became aware of a recent experiment on MoSe$_2$-WSe$_2$ where exciton condensation with high transition temperatures above $100$ K consistent with our predictions were reported.}
\\

This work was partially supported by the Fonds Wetenschappelijk Onderzoek (FWO-Vl), the Methusalem Foundation and the FLAG-ERA project TRANS-2D-TMD.  We thank A. R. Hamilton and A. Vargas-Paredes for useful discussions.

\appendix

\section{Appendix: Mean field equations}
\label{appendix}

To describe our system we introduce the temperature dependent normal and anomalous multiband Matsubara Green functions, with subband indices $\alpha$ and $\beta$, 
\begin{equation}
\begin{cases}
\mathcal{G}^{\alpha\beta}(k,\tau)&=-<T c^{\alpha}_{k}(\tau) c^{\beta \dagger}_{k}(0)> \\
\mathcal{F}^{\alpha\beta}(k,\tau)&=-<T c^{\alpha}_{k}(\tau) d^{\beta}_{k}(0)>.
\end{cases}
\label{eq:GreenFun}
\end{equation} 
The mean field equations for the gaps and the densities are\cite{Shanenko2015}:
\begin{equation}
\Delta_{\alpha\beta}(k)=-\frac{T}{L^2}\sum_{\substack{\alpha',\beta',\\ k', i \omega_n} } F^{\alpha\beta\alpha'\beta'}_{kk'} \, V^{eh}_{k\, k'} \,\mathcal{F}^{\alpha'\beta'}(k', i \omega_n)
\label{eq:gap_sum}
\end{equation}
\begin{equation}
n_{\alpha\beta}=\frac{T}{L^2}\sum_{k, i \omega_n } \mathcal{G}^{\alpha\beta}(k, i \omega_n)
\label{eq:density_sum}
\end{equation}
where $F^{\alpha\beta\alpha'\beta'}_{kk'}= \Braket{\alpha' k'|\alpha k}\Braket{\beta k|\beta' k'}$ is the form factor representing the overlap of the single particle wave functions. On the right hand side of Eq.\ \ref{eq:gap_sum}, the gaps $\Delta_{\alpha\beta}(k)$ appear implicitly in the $\mathcal{F}^{\alpha\beta}$.

Since we are neglecting the cross-pairing contributions, we retain the Green functions and the form factors only for $\alpha=\beta$($\alpha'=\beta'$). 
The screened Coulomb interaction $V^{eh}_{k\, k'}$ conserves the spin of the electron-hole pair and there are no spin-flip scattering processes implying $F^{\beta\beta\beta'\beta'}_{kk'}= 0$ for $\beta\neq\beta'$, so Josephson-like pair transfers are forbidden. The resulting gap equations are thus decoupled.
For brevity, we adopt the notation $F^{\beta\beta\beta'\beta'}_{kk'}\equiv F^{\beta\beta'}_{kk'}$.

In terms of Bogoliubov amplitudes:
\begin{equation}
\!\!\!\!  v^2_{\beta}(k) = \frac{1}{2}\left(1 -\frac{\xi_\beta(k)}{E_\beta(k)} \right);\:
u^2_{\beta}(k) = \frac{1}{2}\left(1 +\frac{\xi_\beta(k)}{E_\beta(k)}\right),
\label{eq:B_a}
\end{equation}
Eqs.\ (\ref{eq:GreenFun}) become
\begin{align}
\mathcal{G}^{\beta\beta}(k, i\omega_n)&=\frac{u_\beta^2}{i\omega_n-E_\beta^-} + \frac{v_\beta^2}{i\omega_n+E_\beta^+} \\
\mathcal{F}^{\beta\beta}(k, i \varepsilon_n)&=\frac{u_\beta v_\beta}{i\omega_n-E_\beta^-} + \frac{u_\beta v_\beta}{i\omega_n+E_\beta^+},
\label{eq:GreenFun_uv}
\end{align}
with $E_\beta^\pm$ defined in the main manuscript.

Performing the summation over the Matsubara frequencies $\omega_n= \pi T (2n + 1)$ in the limit of zero temperature, we obtain the gap equations (Eqs.\ (3-4)) and the density equations (Eqs.\ (7-8)) in the main manuscript. 

The polarizabilities in the presence of the superfluid are \cite{Lozovik2012}:
\begin{equation}
\Pi_n(q,\Omega_l)= T\,\frac{g_s g_v}{L^2}\hspace{-0.2cm}\sum_{\beta,k', i \omega_n }\hspace{-0.2cm}
F^{\beta\beta}_{kk'} \mathcal{G}^{\beta\beta}(k', i\omega_n+i\Omega_l) \mathcal{G}^{\beta\beta}(k, i\omega_n)
\label{Pi_n}
\end{equation}
\begin{equation}
\Pi_a(q,\Omega_l)= T\,\frac{g_s g_v}{L^2}\hspace{-0.2cm}\sum_{\beta,k', i \omega_n }\hspace{-0.2cm}
F^{\beta\beta}_{kk'}\mathcal{F}^{\beta\beta}(k', i\omega_n+i\Omega_l)\mathcal{F}^{\beta\beta}(k, i\omega_n)
\label{Pi_a}
\end{equation}
where $q=|\textbf{k}-\textbf{k}'|$.
The polarizabilities in the effective electron-hole interaction (Eq.\ (5) in the main manuscript) are obtained by evaluating Eqs.\ (\ref{Pi_n}) and (\ref{Pi_a}) 
at zero temperature in the static limit, $\Omega_l\rightarrow0$.


\begin{thebibliography}{41}%
\makeatletter
\providecommand \@ifxundefined [1]{%
 \@ifx{#1\undefined}
}%
\providecommand \@ifnum [1]{%
 \ifnum #1\expandafter \@firstoftwo
 \else \expandafter \@secondoftwo
 \fi
}%
\providecommand \@ifx [1]{%
 \ifx #1\expandafter \@firstoftwo
 \else \expandafter \@secondoftwo
 \fi
}%
\providecommand \natexlab [1]{#1}%
\providecommand \enquote  [1]{``#1''}%
\providecommand \bibnamefont  [1]{#1}%
\providecommand \bibfnamefont [1]{#1}%
\providecommand \citenamefont [1]{#1}%
\providecommand \href@noop [0]{\@secondoftwo}%
\providecommand \href [0]{\begingroup \@sanitize@url \@href}%
\providecommand \@href[1]{\@@startlink{#1}\@@href}%
\providecommand \@@href[1]{\endgroup#1\@@endlink}%
\providecommand \@sanitize@url [0]{\catcode `\\12\catcode `\$12\catcode
  `\&12\catcode `\#12\catcode `\^12\catcode `\_12\catcode `\%12\relax}%
\providecommand \@@startlink[1]{}%
\providecommand \@@endlink[0]{}%
\providecommand \url  [0]{\begingroup\@sanitize@url \@url }%
\providecommand \@url [1]{\endgroup\@href {#1}{\urlprefix }}%
\providecommand \urlprefix  [0]{URL }%
\providecommand \Eprint [0]{\href }%
\providecommand \doibase [0]{http://dx.doi.org/}%
\providecommand \selectlanguage [0]{\@gobble}%
\providecommand \bibinfo  [0]{\@secondoftwo}%
\providecommand \bibfield  [0]{\@secondoftwo}%
\providecommand \translation [1]{[#1]}%
\providecommand \BibitemOpen [0]{}%
\providecommand \bibitemStop [0]{}%
\providecommand \bibitemNoStop [0]{.\EOS\space}%
\providecommand \EOS [0]{\spacefactor3000\relax}%
\providecommand \BibitemShut  [1]{\csname bibitem#1\endcsname}%
\let\auto@bib@innerbib\@empty
\bibitem [{\citenamefont {Burg}\ \emph {et~al.}(2018)\citenamefont {Burg},
  \citenamefont {Prasad}, \citenamefont {Kim}, \citenamefont {Taniguchi},
  \citenamefont {Watanabe}, \citenamefont {MacDonald}, \citenamefont
  {Register},\ and\ \citenamefont {Tutuc}}]{Burg2018}%
  \BibitemOpen
  \bibfield  {author} {\bibinfo {author} {\bibfnamefont {G.~W.}\ \bibnamefont
  {Burg}}, \bibinfo {author} {\bibfnamefont {N.}~\bibnamefont {Prasad}},
  \bibinfo {author} {\bibfnamefont {K.}~\bibnamefont {Kim}}, \bibinfo {author}
  {\bibfnamefont {T.}~\bibnamefont {Taniguchi}}, \bibinfo {author}
  {\bibfnamefont {K.}~\bibnamefont {Watanabe}}, \bibinfo {author}
  {\bibfnamefont {A.~H.}\ \bibnamefont {MacDonald}}, \bibinfo {author}
  {\bibfnamefont {L.~F.}\ \bibnamefont {Register}}, \ and\ \bibinfo {author}
  {\bibfnamefont {E.}~\bibnamefont {Tutuc}},\ }\bibfield  {title} {\enquote
  {\bibinfo {title} {Strongly enhanced tunneling at total charge neutrality in
  double-bilayer graphene-{${\mathrm{WSe}}_{2}$} heterostructures},}\ }\href
  {\doibase 10.1103/PhysRevLett.120.177702} {\bibfield  {journal} {\bibinfo
  {journal} {Phys. Rev. Lett.}\ }\textbf {\bibinfo {volume} {120}},\ \bibinfo
  {pages} {177702} (\bibinfo {year} {2018})}\BibitemShut {NoStop}%
\bibitem [{\citenamefont {Conti}\ \emph {et~al.}(2019)\citenamefont {Conti},
  \citenamefont {Perali}, \citenamefont {Peeters},\ and\ \citenamefont
  {Neilson}}]{Conti2019}%
  \BibitemOpen
  \bibfield  {author} {\bibinfo {author} {\bibfnamefont {S.}~\bibnamefont
  {Conti}}, \bibinfo {author} {\bibfnamefont {A.}~\bibnamefont {Perali}},
  \bibinfo {author} {\bibfnamefont {F.~M.}\ \bibnamefont {Peeters}}, \ and\
  \bibinfo {author} {\bibfnamefont {D.}~\bibnamefont {Neilson}},\ }\bibfield
  {title} {\enquote {\bibinfo {title} {Multicomponent screening and
  superfluidity in gapped electron-hole double bilayer graphene with realistic
  bands},}\ }\href {\doibase 10.1103/PhysRevB.99.144517} {\bibfield  {journal}
  {\bibinfo  {journal} {Phys. Rev. B}\ }\textbf {\bibinfo {volume} {99}},\
  \bibinfo {pages} {144517} (\bibinfo {year} {2019})}\BibitemShut {NoStop}%
\bibitem [{\citenamefont {Zhang}\ \emph {et~al.}(2009)\citenamefont {Zhang},
  \citenamefont {Tang}, \citenamefont {Girit}, \citenamefont {Hao},
  \citenamefont {Martin}, \citenamefont {Zettl}, \citenamefont {Crommie},
  \citenamefont {Shen},\ and\ \citenamefont {Wang}}]{Zhang2009}%
  \BibitemOpen
  \bibfield  {author} {\bibinfo {author} {\bibfnamefont {Y.}~\bibnamefont
  {Zhang}}, \bibinfo {author} {\bibfnamefont {T.~T.}\ \bibnamefont {Tang}},
  \bibinfo {author} {\bibfnamefont {C.}~\bibnamefont {Girit}}, \bibinfo
  {author} {\bibfnamefont {Z.}~\bibnamefont {Hao}}, \bibinfo {author}
  {\bibfnamefont {M.~C.}\ \bibnamefont {Martin}}, \bibinfo {author}
  {\bibfnamefont {A.}~\bibnamefont {Zettl}}, \bibinfo {author} {\bibfnamefont
  {M.~F.}\ \bibnamefont {Crommie}}, \bibinfo {author} {\bibfnamefont {Y.~R.}\
  \bibnamefont {Shen}}, \ and\ \bibinfo {author} {\bibfnamefont
  {F.}~\bibnamefont {Wang}},\ }\bibfield  {title} {\enquote {\bibinfo {title}
  {Direct observation of a widely tunable bandgap in bilayer graphene},}\
  }\href {https://www.nature.com/articles/nature08105} {\bibfield  {journal}
  {\bibinfo  {journal} {Nature (London)}\ }\textbf {\bibinfo {volume} {459}},\
  \bibinfo {pages} {820} (\bibinfo {year} {2009})}\BibitemShut {NoStop}%
\bibitem [{\citenamefont {Mak}\ \emph {et~al.}(2010)\citenamefont {Mak},
  \citenamefont {Lee}, \citenamefont {Hone}, \citenamefont {Shan},\ and\
  \citenamefont {Heinz}}]{Mak2010}%
  \BibitemOpen
  \bibfield  {author} {\bibinfo {author} {\bibfnamefont {K.~F.}\ \bibnamefont
  {Mak}}, \bibinfo {author} {\bibfnamefont {C.}~\bibnamefont {Lee}}, \bibinfo
  {author} {\bibfnamefont {J.}~\bibnamefont {Hone}}, \bibinfo {author}
  {\bibfnamefont {J.}~\bibnamefont {Shan}}, \ and\ \bibinfo {author}
  {\bibfnamefont {T.~F.}\ \bibnamefont {Heinz}},\ }\bibfield  {title} {\enquote
  {\bibinfo {title} {Atomically thin {M}o{S}$_2$: a new direct-gap
  semiconductor},}\ }\href
  {https://journals.aps.org/prl/abstract/10.1103/PhysRevLett.105.136805}
  {\bibfield  {journal} {\bibinfo  {journal} {Phys. Rev. Lett.}\ }\textbf
  {\bibinfo {volume} {105}},\ \bibinfo {pages} {136805} (\bibinfo {year}
  {2010})}\BibitemShut {NoStop}%
\bibitem [{\citenamefont {Jiang}(2012)}]{Jiang2012}%
  \BibitemOpen
  \bibfield  {author} {\bibinfo {author} {\bibfnamefont {H.}~\bibnamefont
  {Jiang}},\ }\bibfield  {title} {\enquote {\bibinfo {title} {Electronic band
  structures of molybdenum and tungsten dichalcogenides by the {GW}
  approach},}\ }\href {https://pubs.acs.org/doi/abs/10.1021/jp300079d}
  {\bibfield  {journal} {\bibinfo  {journal} {J. Phys. Chem. C}\ }\textbf
  {\bibinfo {volume} {116}},\ \bibinfo {pages} {7664} (\bibinfo {year}
  {2012})}\BibitemShut {NoStop}%
\bibitem [{\citenamefont {Fogler}\ \emph {et~al.}(2014)\citenamefont {Fogler},
  \citenamefont {Butov},\ and\ \citenamefont {Novoselov}}]{Fogler2014}%
  \BibitemOpen
  \bibfield  {author} {\bibinfo {author} {\bibfnamefont {M.~M.}\ \bibnamefont
  {Fogler}}, \bibinfo {author} {\bibfnamefont {L.~V.}\ \bibnamefont {Butov}}, \
  and\ \bibinfo {author} {\bibfnamefont {K.~S.}\ \bibnamefont {Novoselov}},\
  }\bibfield  {title} {\enquote {\bibinfo {title} {High-temperature
  superfluidity with indirect excitons in van der {W}aals heterostructures},}\
  }\href {https://www.nature.com/articles/ncomms5555} {\bibfield  {journal}
  {\bibinfo  {journal} {Nat. Commun.}\ }\textbf {\bibinfo {volume} {5}},\
  \bibinfo {pages} {4555} (\bibinfo {year} {2014})}\BibitemShut {NoStop}%
\bibitem [{\citenamefont {Rivera}\ \emph {et~al.}(2015)\citenamefont {Rivera},
  \citenamefont {Schaibley}, \citenamefont {Jones}, \citenamefont {Ross},
  \citenamefont {Wu}, \citenamefont {Aivazian}, \citenamefont {Klement},
  \citenamefont {Seyler}, \citenamefont {Clark}, \citenamefont {Ghimire} \emph
  {et~al.}}]{Rivera2015}%
  \BibitemOpen
  \bibfield  {author} {\bibinfo {author} {\bibfnamefont {P.}~\bibnamefont
  {Rivera}}, \bibinfo {author} {\bibfnamefont {J.~R.}\ \bibnamefont
  {Schaibley}}, \bibinfo {author} {\bibfnamefont {A.~M.}\ \bibnamefont
  {Jones}}, \bibinfo {author} {\bibfnamefont {J.~S.}\ \bibnamefont {Ross}},
  \bibinfo {author} {\bibfnamefont {S.}~\bibnamefont {Wu}}, \bibinfo {author}
  {\bibfnamefont {G.}~\bibnamefont {Aivazian}}, \bibinfo {author}
  {\bibfnamefont {P.}~\bibnamefont {Klement}}, \bibinfo {author} {\bibfnamefont
  {K.}~\bibnamefont {Seyler}}, \bibinfo {author} {\bibfnamefont
  {G.}~\bibnamefont {Clark}}, \bibinfo {author} {\bibfnamefont {N.~J.}\
  \bibnamefont {Ghimire}},  \emph {et~al.},\ }\bibfield  {title} {\enquote
  {\bibinfo {title} {Observation of long-lived interlayer excitons in monolayer
  {M}o{S}e$_2$-{WS}e$_2$ heterostructures},}\ }\href
  {https://www.nature.com/articles/ncomms7242} {\bibfield  {journal} {\bibinfo
  {journal} {Nat. Commun.}\ }\textbf {\bibinfo {volume} {6}},\ \bibinfo {pages}
  {6242} (\bibinfo {year} {2015})}\BibitemShut {NoStop}%
\bibitem [{\citenamefont {Ovesen}\ \emph {et~al.}(2019)\citenamefont {Ovesen},
  \citenamefont {Brem}, \citenamefont {Linder{\"a}lv}, \citenamefont {Kuisma},
  \citenamefont {Korn}, \citenamefont {Erhart}, \citenamefont {Selig},\ and\
  \citenamefont {Malic}}]{Ovesen2019}%
  \BibitemOpen
  \bibfield  {author} {\bibinfo {author} {\bibfnamefont {S.}~\bibnamefont
  {Ovesen}}, \bibinfo {author} {\bibfnamefont {S.}~\bibnamefont {Brem}},
  \bibinfo {author} {\bibfnamefont {C.}~\bibnamefont {Linder{\"a}lv}}, \bibinfo
  {author} {\bibfnamefont {M.}~\bibnamefont {Kuisma}}, \bibinfo {author}
  {\bibfnamefont {T.}~\bibnamefont {Korn}}, \bibinfo {author} {\bibfnamefont
  {P.}~\bibnamefont {Erhart}}, \bibinfo {author} {\bibfnamefont
  {M.}~\bibnamefont {Selig}}, \ and\ \bibinfo {author} {\bibfnamefont
  {E.}~\bibnamefont {Malic}},\ }\bibfield  {title} {\enquote {\bibinfo {title}
  {Interlayer exciton dynamics in van der waals heterostructures},}\ }\href
  {https://www.nature.com/articles/s42005-019-0122-z} {\bibfield  {journal}
  {\bibinfo  {journal} {Communications Physics}\ }\textbf {\bibinfo {volume}
  {2}},\ \bibinfo {pages} {23} (\bibinfo {year} {2019})}\BibitemShut {NoStop}%
\bibitem [{\citenamefont {F{\"o}rg}\ \emph {et~al.}(2019)\citenamefont
  {F{\"o}rg}, \citenamefont {Colombier}, \citenamefont {Patel}, \citenamefont
  {Lindlau}, \citenamefont {Mohite}, \citenamefont {Yamaguchi}, \citenamefont
  {Glazov}, \citenamefont {Hunger},\ and\ \citenamefont
  {H{\"o}gele}}]{Forg2019}%
  \BibitemOpen
  \bibfield  {author} {\bibinfo {author} {\bibfnamefont {M.}~\bibnamefont
  {F{\"o}rg}}, \bibinfo {author} {\bibfnamefont {L.}~\bibnamefont {Colombier}},
  \bibinfo {author} {\bibfnamefont {R.~K.}\ \bibnamefont {Patel}}, \bibinfo
  {author} {\bibfnamefont {J.}~\bibnamefont {Lindlau}}, \bibinfo {author}
  {\bibfnamefont {A.~D}\ \bibnamefont {Mohite}}, \bibinfo {author}
  {\bibfnamefont {H.}~\bibnamefont {Yamaguchi}}, \bibinfo {author}
  {\bibfnamefont {M.~M.}\ \bibnamefont {Glazov}}, \bibinfo {author}
  {\bibfnamefont {D.}~\bibnamefont {Hunger}}, \ and\ \bibinfo {author}
  {\bibfnamefont {A.}~\bibnamefont {H{\"o}gele}},\ }\bibfield  {title}
  {\enquote {\bibinfo {title} {Cavity-control of interlayer excitons in van der
  {W}aals heterostructures},}\ }\href
  {https://www.nature.com/articles/s41467-019-11620-z} {\bibfield  {journal}
  {\bibinfo  {journal} {Nat. Commun.}\ }\textbf {\bibinfo {volume} {10}},\
  \bibinfo {pages} {3697} (\bibinfo {year} {2019})}\BibitemShut {NoStop}%
\bibitem [{\citenamefont {Britnell}\ \emph {et~al.}(2012)\citenamefont
  {Britnell}, \citenamefont {Gorbachev}, \citenamefont {Jalil}, \citenamefont
  {Belle}, \citenamefont {Schedin}, \citenamefont {Katsnelson}, \citenamefont
  {Eaves}, \citenamefont {Morozov}, \citenamefont {Mayorov}, \citenamefont
  {Peres}, \citenamefont {Castro~Neto}, \citenamefont {Leist}, \citenamefont
  {Geim}, \citenamefont {Ponomarenko},\ and\ \citenamefont
  {Novoselov}}]{Britnell2012a}%
  \BibitemOpen
  \bibfield  {author} {\bibinfo {author} {\bibfnamefont {L.}~\bibnamefont
  {Britnell}}, \bibinfo {author} {\bibfnamefont {R.~V.}\ \bibnamefont
  {Gorbachev}}, \bibinfo {author} {\bibfnamefont {R.}~\bibnamefont {Jalil}},
  \bibinfo {author} {\bibfnamefont {B.~D.}\ \bibnamefont {Belle}}, \bibinfo
  {author} {\bibfnamefont {F.}~\bibnamefont {Schedin}}, \bibinfo {author}
  {\bibfnamefont {M.~I.}\ \bibnamefont {Katsnelson}}, \bibinfo {author}
  {\bibfnamefont {L.}~\bibnamefont {Eaves}}, \bibinfo {author} {\bibfnamefont
  {S.~V.}\ \bibnamefont {Morozov}}, \bibinfo {author} {\bibfnamefont {A.~S.}\
  \bibnamefont {Mayorov}}, \bibinfo {author} {\bibfnamefont {N.~MR.}\
  \bibnamefont {Peres}}, \bibinfo {author} {\bibfnamefont {A.~H.}\ \bibnamefont
  {Castro~Neto}}, \bibinfo {author} {\bibfnamefont {J.}~\bibnamefont {Leist}},
  \bibinfo {author} {\bibfnamefont {A.~K.}\ \bibnamefont {Geim}}, \bibinfo
  {author} {\bibfnamefont {L.~A.}\ \bibnamefont {Ponomarenko}}, \ and\ \bibinfo
  {author} {\bibfnamefont {K.~S.}\ \bibnamefont {Novoselov}},\ }\bibfield
  {title} {\enquote {\bibinfo {title} {Electron tunneling through ultrathin
  {B}oron {N}itride crystalline barriers},}\ }\href
  {https://pubs.acs.org/doi/full/10.1021/nl3002205} {\bibfield  {journal}
  {\bibinfo  {journal} {Nano Lett.}\ }\textbf {\bibinfo {volume} {12}},\
  \bibinfo {pages} {1707} (\bibinfo {year} {2012})}\BibitemShut {NoStop}%
\bibitem [{\citenamefont {Bianconi}(2013)}]{Bianconi2013}%
  \BibitemOpen
  \bibfield  {author} {\bibinfo {author} {\bibfnamefont {A.}~\bibnamefont
  {Bianconi}},\ }\bibfield  {title} {\enquote {\bibinfo {title} {Quantum
  materials: Shape resonances in superstripes},}\ }\href
  {https://www.nature.com/articles/nphys2738} {\bibfield  {journal} {\bibinfo
  {journal} {Nat. Phys.}\ }\textbf {\bibinfo {volume} {9}},\ \bibinfo {pages}
  {536} (\bibinfo {year} {2013})}\BibitemShut {NoStop}%
\bibitem [{\citenamefont {Shanenko}\ \emph {et~al.}(2012)\citenamefont
  {Shanenko}, \citenamefont {Croitoru}, \citenamefont {Vagov}, \citenamefont
  {Axt}, \citenamefont {Perali},\ and\ \citenamefont {Peeters}}]{Shanenko2012}%
  \BibitemOpen
  \bibfield  {author} {\bibinfo {author} {\bibfnamefont {A.~A.}\ \bibnamefont
  {Shanenko}}, \bibinfo {author} {\bibfnamefont {M.~D.}\ \bibnamefont
  {Croitoru}}, \bibinfo {author} {\bibfnamefont {A.~V.}\ \bibnamefont {Vagov}},
  \bibinfo {author} {\bibfnamefont {V.~M.}\ \bibnamefont {Axt}}, \bibinfo
  {author} {\bibfnamefont {A.}~\bibnamefont {Perali}}, \ and\ \bibinfo {author}
  {\bibfnamefont {F.~M.}\ \bibnamefont {Peeters}},\ }\bibfield  {title}
  {\enquote {\bibinfo {title} {Atypical {BCS-BEC} crossover induced by
  quantum-size effects},}\ }\href
  {https://journals.aps.org/pra/abstract/10.1103/PhysRevA.86.033612} {\bibfield
   {journal} {\bibinfo  {journal} {Phys. Rev. A}\ }\textbf {\bibinfo {volume}
  {86}},\ \bibinfo {pages} {033612} (\bibinfo {year} {2012})}\BibitemShut
  {NoStop}%
\bibitem [{\citenamefont {Mizohata}\ \emph {et~al.}(2013)\citenamefont
  {Mizohata}, \citenamefont {Ichioka},\ and\ \citenamefont
  {Machida}}]{Mizohata2013}%
  \BibitemOpen
  \bibfield  {author} {\bibinfo {author} {\bibfnamefont {Y.}~\bibnamefont
  {Mizohata}}, \bibinfo {author} {\bibfnamefont {M.}~\bibnamefont {Ichioka}}, \
  and\ \bibinfo {author} {\bibfnamefont {K.}~\bibnamefont {Machida}},\
  }\bibfield  {title} {\enquote {\bibinfo {title} {Multiple-gap structure in
  electric-field-induced surface superconductivity},}\ }\href
  {https://journals.aps.org/prb/abstract/10.1103/PhysRevB.87.014505} {\bibfield
   {journal} {\bibinfo  {journal} {Phys. Rev. B}\ }\textbf {\bibinfo {volume}
  {87}},\ \bibinfo {pages} {014505} (\bibinfo {year} {2013})}\BibitemShut
  {NoStop}%
\bibitem [{\citenamefont {Singh}\ \emph {et~al.}(2019)\citenamefont {Singh},
  \citenamefont {Jouan}, \citenamefont {Herranz}, \citenamefont {Scigaj},
  \citenamefont {S{\'a}nchez}, \citenamefont {Benfatto}, \citenamefont
  {Caprara}, \citenamefont {Grilli}, \citenamefont {Saiz}, \citenamefont
  {Cou{\"e}do}, \citenamefont {Feuillet-Palma}, \citenamefont {Lesueur},\ and\
  \citenamefont {Bergeal}}]{Singh2019}%
  \BibitemOpen
  \bibfield  {author} {\bibinfo {author} {\bibfnamefont {G.}~\bibnamefont
  {Singh}}, \bibinfo {author} {\bibfnamefont {A.}~\bibnamefont {Jouan}},
  \bibinfo {author} {\bibfnamefont {G.}~\bibnamefont {Herranz}}, \bibinfo
  {author} {\bibfnamefont {M.}~\bibnamefont {Scigaj}}, \bibinfo {author}
  {\bibfnamefont {F.}~\bibnamefont {S{\'a}nchez}}, \bibinfo {author}
  {\bibfnamefont {L.}~\bibnamefont {Benfatto}}, \bibinfo {author}
  {\bibfnamefont {S.}~\bibnamefont {Caprara}}, \bibinfo {author} {\bibfnamefont
  {M.}~\bibnamefont {Grilli}}, \bibinfo {author} {\bibfnamefont
  {G.}~\bibnamefont {Saiz}}, \bibinfo {author} {\bibfnamefont {F.}~\bibnamefont
  {Cou{\"e}do}}, \bibinfo {author} {\bibfnamefont {C.}~\bibnamefont
  {Feuillet-Palma}}, \bibinfo {author} {\bibfnamefont {J.}~\bibnamefont
  {Lesueur}}, \ and\ \bibinfo {author} {\bibfnamefont {N.}~\bibnamefont
  {Bergeal}},\ }\bibfield  {title} {\enquote {\bibinfo {title} {Gap suppression
  at a {L}ifshitz transition in a multi-condensate superconductor},}\ }\href
  {\doibase 10.1038/s41563-019-0354-z} {\bibfield  {journal} {\bibinfo
  {journal} {Nature Materials}\ }\textbf {\bibinfo {volume} {18}},\ \bibinfo
  {pages} {948} (\bibinfo {year} {2019})}\BibitemShut {NoStop}%
\bibitem [{\citenamefont {Xiao}\ \emph {et~al.}(2012)\citenamefont {Xiao},
  \citenamefont {Liu}, \citenamefont {Feng}, \citenamefont {Xu},\ and\
  \citenamefont {Yao}}]{Xiao2012}%
  \BibitemOpen
  \bibfield  {author} {\bibinfo {author} {\bibfnamefont {D.}~\bibnamefont
  {Xiao}}, \bibinfo {author} {\bibfnamefont {G.~B.}\ \bibnamefont {Liu}},
  \bibinfo {author} {\bibfnamefont {W.}~\bibnamefont {Feng}}, \bibinfo {author}
  {\bibfnamefont {X.}~\bibnamefont {Xu}}, \ and\ \bibinfo {author}
  {\bibfnamefont {W.}~\bibnamefont {Yao}},\ }\bibfield  {title} {\enquote
  {\bibinfo {title} {Coupled spin and valley physics in monolayers of {MoS}$_2$
  and other group-{VI} dichalcogenides},}\ }\href {\doibase
  10.1103/PhysRevLett.108.196802} {\bibfield  {journal} {\bibinfo  {journal}
  {Phys. Rev. Lett.}\ }\textbf {\bibinfo {volume} {108}},\ \bibinfo {pages}
  {196802} (\bibinfo {year} {2012})}\BibitemShut {NoStop}%
\bibitem [{\citenamefont {Zhu}\ \emph {et~al.}(2011)\citenamefont {Zhu},
  \citenamefont {Cheng},\ and\ \citenamefont {Schwingenschl\"ogl}}]{Zhu2011}%
  \BibitemOpen
  \bibfield  {author} {\bibinfo {author} {\bibfnamefont {Z.~Y.}\ \bibnamefont
  {Zhu}}, \bibinfo {author} {\bibfnamefont {Y.~C.}\ \bibnamefont {Cheng}}, \
  and\ \bibinfo {author} {\bibfnamefont {U.}~\bibnamefont
  {Schwingenschl\"ogl}},\ }\bibfield  {title} {\enquote {\bibinfo {title}
  {Giant spin-orbit-induced spin splitting in two-dimensional transition-metal
  dichalcogenide semiconductors},}\ }\href {\doibase
  10.1103/PhysRevB.84.153402} {\bibfield  {journal} {\bibinfo  {journal} {Phys.
  Rev. B}\ }\textbf {\bibinfo {volume} {84}},\ \bibinfo {pages} {153402}
  (\bibinfo {year} {2011})}\BibitemShut {NoStop}%
\bibitem [{\citenamefont {Ko\'smider}\ \emph {et~al.}(2013)\citenamefont
  {Ko\'smider}, \citenamefont {Gonz\'alez},\ and\ \citenamefont
  {Fern\'andez-Rossier}}]{Kosmider2013}%
  \BibitemOpen
  \bibfield  {author} {\bibinfo {author} {\bibfnamefont {K.}~\bibnamefont
  {Ko\'smider}}, \bibinfo {author} {\bibfnamefont {J.~W.}\ \bibnamefont
  {Gonz\'alez}}, \ and\ \bibinfo {author} {\bibfnamefont {J.}~\bibnamefont
  {Fern\'andez-Rossier}},\ }\bibfield  {title} {\enquote {\bibinfo {title}
  {Large spin splitting in the conduction band of transition metal
  dichalcogenide monolayers},}\ }\href {\doibase 10.1103/PhysRevB.88.245436}
  {\bibfield  {journal} {\bibinfo  {journal} {Phys. Rev. B}\ }\textbf {\bibinfo
  {volume} {88}},\ \bibinfo {pages} {245436} (\bibinfo {year}
  {2013})}\BibitemShut {NoStop}%
\bibitem [{\citenamefont {Conti}\ \emph {et~al.}(2017)\citenamefont {Conti},
  \citenamefont {Perali}, \citenamefont {Peeters},\ and\ \citenamefont
  {Neilson}}]{Conti2017}%
  \BibitemOpen
  \bibfield  {author} {\bibinfo {author} {\bibfnamefont {S.}~\bibnamefont
  {Conti}}, \bibinfo {author} {\bibfnamefont {A.}~\bibnamefont {Perali}},
  \bibinfo {author} {\bibfnamefont {F.~M.}\ \bibnamefont {Peeters}}, \ and\
  \bibinfo {author} {\bibfnamefont {D.}~\bibnamefont {Neilson}},\ }\bibfield
  {title} {\enquote {\bibinfo {title} {Multicomponent electron-hole
  superfluidity and the {BCS-BEC} crossover in double bilayer graphene},}\
  }\href {\doibase https://doi.org/10.1103/PhysRevLett.119.257002} {\bibfield
  {journal} {\bibinfo  {journal} {Phys. Rev. Lett.}\ }\textbf {\bibinfo
  {volume} {119}},\ \bibinfo {pages} {257002} (\bibinfo {year}
  {2017})}\BibitemShut {NoStop}%
\bibitem [{\citenamefont {Van~der Donck}\ and\ \citenamefont
  {Peeters}(2018)}]{Vanderdonck2018}%
  \BibitemOpen
  \bibfield  {author} {\bibinfo {author} {\bibfnamefont {M.}~\bibnamefont
  {Van~der Donck}}\ and\ \bibinfo {author} {\bibfnamefont {F.~M.}\ \bibnamefont
  {Peeters}},\ }\bibfield  {title} {\enquote {\bibinfo {title} {Interlayer
  excitons in transition metal dichalcogenide heterostructures},}\ }\href
  {https://journals.aps.org/prb/abstract/10.1103/PhysRevB.98.115104} {\bibfield
   {journal} {\bibinfo  {journal} {Phys. Rev. B}\ }\textbf {\bibinfo {volume}
  {98}},\ \bibinfo {pages} {115104} (\bibinfo {year} {2018})}\BibitemShut
  {NoStop}%
\bibitem [{\citenamefont {Shanenko}\ \emph {et~al.}(2015)\citenamefont
  {Shanenko}, \citenamefont {Aguiar}, \citenamefont {Vagov}, \citenamefont
  {Croitoru},\ and\ \citenamefont {Milo{\v{s}}evi{\'c}}}]{Shanenko2015}%
  \BibitemOpen
  \bibfield  {author} {\bibinfo {author} {\bibfnamefont {A.~A.}\ \bibnamefont
  {Shanenko}}, \bibinfo {author} {\bibfnamefont {J.~A.}\ \bibnamefont
  {Aguiar}}, \bibinfo {author} {\bibfnamefont {A.}~\bibnamefont {Vagov}},
  \bibinfo {author} {\bibfnamefont {M.~D.}\ \bibnamefont {Croitoru}}, \ and\
  \bibinfo {author} {\bibfnamefont {M.~V.}\ \bibnamefont
  {Milo{\v{s}}evi{\'c}}},\ }\bibfield  {title} {\enquote {\bibinfo {title}
  {Atomically flat superconducting nanofilms: multiband properties and
  mean-field theory},}\ }\href
  {https://iopscience.iop.org/article/10.1088/0953-2048/28/5/054001} {\bibfield
   {journal} {\bibinfo  {journal} {Supercond. Sci. Tech.}\ }\textbf {\bibinfo
  {volume} {28}},\ \bibinfo {pages} {054001} (\bibinfo {year}
  {2015})}\BibitemShut {NoStop}%
\bibitem [{\citenamefont {Lozovik}\ and\ \citenamefont
  {Sokolik}(2009)}]{Lozovik2009}%
  \BibitemOpen
  \bibfield  {author} {\bibinfo {author} {\bibfnamefont {Y.~E}\ \bibnamefont
  {Lozovik}}\ and\ \bibinfo {author} {\bibfnamefont {A.~A.}\ \bibnamefont
  {Sokolik}},\ }\bibfield  {title} {\enquote {\bibinfo {title} {Multi-band
  pairing of ultrarelativistic electrons and holes in graphene bilayer},}\
  }\href {https://www.sciencedirect.com/science/article/pii/S0375960109013413}
  {\bibfield  {journal} {\bibinfo  {journal} {Phys. Rev. A}\ }\textbf {\bibinfo
  {volume} {374}},\ \bibinfo {pages} {326} (\bibinfo {year}
  {2009})}\BibitemShut {NoStop}%
\bibitem [{\citenamefont {Lozovik}\ \emph {et~al.}(2012)\citenamefont
  {Lozovik}, \citenamefont {Ogarkov},\ and\ \citenamefont
  {Sokolik}}]{Lozovik2012}%
  \BibitemOpen
  \bibfield  {author} {\bibinfo {author} {\bibfnamefont {Y.~E.}\ \bibnamefont
  {Lozovik}}, \bibinfo {author} {\bibfnamefont {S.~L.}\ \bibnamefont
  {Ogarkov}}, \ and\ \bibinfo {author} {\bibfnamefont {A.~A.}\ \bibnamefont
  {Sokolik}},\ }\bibfield  {title} {\enquote {\bibinfo {title} {Condensation of
  electron-hole pairs in a two-layer graphene system: {C}orrelation effects},}\
  }\href {https://journals.aps.org/prb/abstract/10.1103/PhysRevB.86.045429}
  {\bibfield  {journal} {\bibinfo  {journal} {Phys. Rev. B}\ }\textbf {\bibinfo
  {volume} {86}},\ \bibinfo {pages} {045429} (\bibinfo {year}
  {2012})}\BibitemShut {NoStop}%
\bibitem [{\citenamefont {Perali}\ \emph {et~al.}(2013)\citenamefont {Perali},
  \citenamefont {Neilson},\ and\ \citenamefont {Hamilton}}]{Perali2013}%
  \BibitemOpen
  \bibfield  {author} {\bibinfo {author} {\bibfnamefont {A.}~\bibnamefont
  {Perali}}, \bibinfo {author} {\bibfnamefont {D.}~\bibnamefont {Neilson}}, \
  and\ \bibinfo {author} {\bibfnamefont {A.~R.}\ \bibnamefont {Hamilton}},\
  }\bibfield  {title} {\enquote {\bibinfo {title} {High-temperature
  superfluidity in double-bilayer graphene},}\ }\href {\doibase
  10.1103/PhysRevLett.110.146803} {\bibfield  {journal} {\bibinfo  {journal}
  {Phys. Rev. Lett.}\ }\textbf {\bibinfo {volume} {110}},\ \bibinfo {pages}
  {146803} (\bibinfo {year} {2013})}\BibitemShut {NoStop}%
\bibitem [{\citenamefont {Neilson}\ \emph {et~al.}(2014)\citenamefont
  {Neilson}, \citenamefont {Perali},\ and\ \citenamefont
  {Hamilton}}]{Neilson2014}%
  \BibitemOpen
  \bibfield  {author} {\bibinfo {author} {\bibfnamefont {D.}~\bibnamefont
  {Neilson}}, \bibinfo {author} {\bibfnamefont {A.}~\bibnamefont {Perali}}, \
  and\ \bibinfo {author} {\bibfnamefont {A.~R.}\ \bibnamefont {Hamilton}},\
  }\bibfield  {title} {\enquote {\bibinfo {title} {Excitonic superfluidity and
  screening in electron-hole bilayer systems},}\ }\href {\doibase
  10.1103/PhysRevB.89.060502} {\bibfield  {journal} {\bibinfo  {journal} {Phys.
  Rev. B}\ }\textbf {\bibinfo {volume} {89}},\ \bibinfo {pages} {060502}
  (\bibinfo {year} {2014})}\BibitemShut {NoStop}%
\bibitem [{\citenamefont {Kochorbe}\ and\ \citenamefont
  {Palistrant}(1993)}]{Kochorbe1993}%
  \BibitemOpen
  \bibfield  {author} {\bibinfo {author} {\bibfnamefont {F.~G.}\ \bibnamefont
  {Kochorbe}}\ and\ \bibinfo {author} {\bibfnamefont {M.~E.}\ \bibnamefont
  {Palistrant}},\ }\bibfield  {title} {\enquote {\bibinfo {title}
  {Superconductivity in a two-band system with low carrier density},}\ }\href
  {https://inis.iaea.org/search/search.aspx?orig_q=RN:25033132} {\bibfield
  {journal} {\bibinfo  {journal} {J. Exp. Theor. Phys.}\ }\textbf {\bibinfo
  {volume} {77}},\ \bibinfo {pages} {442} (\bibinfo {year} {1993})}\BibitemShut
  {NoStop}%
\bibitem [{\citenamefont {Salasnich}\ \emph {et~al.}(2005)\citenamefont
  {Salasnich}, \citenamefont {Manini},\ and\ \citenamefont
  {Parola}}]{Salasnich2005}%
  \BibitemOpen
  \bibfield  {author} {\bibinfo {author} {\bibfnamefont {L.}~\bibnamefont
  {Salasnich}}, \bibinfo {author} {\bibfnamefont {N.}~\bibnamefont {Manini}}, \
  and\ \bibinfo {author} {\bibfnamefont {A.}~\bibnamefont {Parola}},\
  }\bibfield  {title} {\enquote {\bibinfo {title} {Condensate fraction of a
  {F}ermi gas in the {BCS-BEC} crossover},}\ }\href {\doibase
  10.1103/PhysRevA.72.023621} {\bibfield  {journal} {\bibinfo  {journal} {Phys.
  Rev. A}\ }\textbf {\bibinfo {volume} {72}},\ \bibinfo {pages} {023621}
  (\bibinfo {year} {2005})}\BibitemShut {NoStop}%
\bibitem [{\citenamefont {L\'opez~R\'{\i}os}\ \emph {et~al.}(2018)\citenamefont
  {L\'opez~R\'{\i}os}, \citenamefont {Perali}, \citenamefont {Needs},\ and\
  \citenamefont {Neilson}}]{LopezRios2018}%
  \BibitemOpen
  \bibfield  {author} {\bibinfo {author} {\bibfnamefont {P.}~\bibnamefont
  {L\'opez~R\'{\i}os}}, \bibinfo {author} {\bibfnamefont {A.}~\bibnamefont
  {Perali}}, \bibinfo {author} {\bibfnamefont {R.~J.}\ \bibnamefont {Needs}}, \
  and\ \bibinfo {author} {\bibfnamefont {D.}~\bibnamefont {Neilson}},\
  }\bibfield  {title} {\enquote {\bibinfo {title} {Evidence from quantum
  {M}onte {C}arlo simulations of large-gap superfluidity and {BCS}-{BEC}
  crossover in double electron-hole layers},}\ }\href {\doibase
  10.1103/PhysRevLett.120.177701} {\bibfield  {journal} {\bibinfo  {journal}
  {Phys. Rev. Lett.}\ }\textbf {\bibinfo {volume} {120}},\ \bibinfo {pages}
  {177701} (\bibinfo {year} {2018})}\BibitemShut {NoStop}%
\bibitem [{\citenamefont {Kumar}\ \emph {et~al.}(2016)\citenamefont {Kumar},
  \citenamefont {Chauhan}, \citenamefont {Agarwal},\ and\ \citenamefont
  {Bhowmick}}]{Kumar2016}%
  \BibitemOpen
  \bibfield  {author} {\bibinfo {author} {\bibfnamefont {P.}~\bibnamefont
  {Kumar}}, \bibinfo {author} {\bibfnamefont {Y.~S.}\ \bibnamefont {Chauhan}},
  \bibinfo {author} {\bibfnamefont {A.}~\bibnamefont {Agarwal}}, \ and\
  \bibinfo {author} {\bibfnamefont {S.}~\bibnamefont {Bhowmick}},\ }\bibfield
  {title} {\enquote {\bibinfo {title} {Thickness and stacking dependent
  polarizability and dielectric constant of graphene--hexagonal boron nitride
  composite stacks},}\ }\href
  {https://pubs.acs.org/doi/abs/10.1021/acs.jpcc.6b05805} {\bibfield  {journal}
  {\bibinfo  {journal} {J. Phys. Chem. C}\ }\textbf {\bibinfo {volume} {120}},\
  \bibinfo {pages} {17620} (\bibinfo {year} {2016})}\BibitemShut {NoStop}%
\bibitem [{\citenamefont {Randeria}\ \emph {et~al.}(1990)\citenamefont
  {Randeria}, \citenamefont {Duan},\ and\ \citenamefont
  {Shieh}}]{Randeria1990}%
  \BibitemOpen
  \bibfield  {author} {\bibinfo {author} {\bibfnamefont {M.}~\bibnamefont
  {Randeria}}, \bibinfo {author} {\bibfnamefont {J.-M.}\ \bibnamefont {Duan}},
  \ and\ \bibinfo {author} {\bibfnamefont {L.-Y.}\ \bibnamefont {Shieh}},\
  }\bibfield  {title} {\enquote {\bibinfo {title} {Superconductivity in a
  two-dimensional {F}ermi gas: {E}volution from {C}ooper pairing to {B}ose
  condensation},}\ }\href
  {https://journals.aps.org/prb/abstract/10.1103/PhysRevB.41.327} {\bibfield
  {journal} {\bibinfo  {journal} {Phys. Rev. B}\ }\textbf {\bibinfo {volume}
  {41}},\ \bibinfo {pages} {327} (\bibinfo {year} {1990})}\BibitemShut
  {NoStop}%
\bibitem [{\citenamefont {Pistolesi}\ and\ \citenamefont
  {Strinati}(1994)}]{Pistolesi1994}%
  \BibitemOpen
  \bibfield  {author} {\bibinfo {author} {\bibfnamefont {F.}~\bibnamefont
  {Pistolesi}}\ and\ \bibinfo {author} {\bibfnamefont {G.~C.}\ \bibnamefont
  {Strinati}},\ }\bibfield  {title} {\enquote {\bibinfo {title} {Evolution from
  {BCS} superconductivity to {B}ose condensation: Role of the parameter $k_{F}
  \xi$},}\ }\href {\doibase 10.1103/PhysRevB.49.6356} {\bibfield  {journal}
  {\bibinfo  {journal} {Phys. Rev. B}\ }\textbf {\bibinfo {volume} {49}},\
  \bibinfo {pages} {6356} (\bibinfo {year} {1994})}\BibitemShut {NoStop}%
\bibitem [{\citenamefont {Saberi-Pouya}\ \emph {et~al.}(2018)\citenamefont
  {Saberi-Pouya}, \citenamefont {Zarenia}, \citenamefont {Perali},
  \citenamefont {Vazifehshenas},\ and\ \citenamefont {Peeters}}]{Saberi2018}%
  \BibitemOpen
  \bibfield  {author} {\bibinfo {author} {\bibfnamefont {S.}~\bibnamefont
  {Saberi-Pouya}}, \bibinfo {author} {\bibfnamefont {M.}~\bibnamefont
  {Zarenia}}, \bibinfo {author} {\bibfnamefont {A.}~\bibnamefont {Perali}},
  \bibinfo {author} {\bibfnamefont {T.}~\bibnamefont {Vazifehshenas}}, \ and\
  \bibinfo {author} {\bibfnamefont {F.~M.}\ \bibnamefont {Peeters}},\
  }\bibfield  {title} {\enquote {\bibinfo {title} {High-temperature
  electron-hole superfluidity with strong anisotropic gaps in double
  phosphorene monolayers},}\ }\href
  {https://journals.aps.org/prb/abstract/10.1103/PhysRevB.97.174503} {\bibfield
   {journal} {\bibinfo  {journal} {Phys. Rev. B}\ }\textbf {\bibinfo {volume}
  {97}},\ \bibinfo {pages} {174503} (\bibinfo {year} {2018})}\BibitemShut
  {NoStop}%
\bibitem [{\citenamefont {Kosterlitz}\ and\ \citenamefont
  {Thouless}(1973)}]{Kosterlitz1973}%
  \BibitemOpen
  \bibfield  {author} {\bibinfo {author} {\bibfnamefont {J.~M.}\ \bibnamefont
  {Kosterlitz}}\ and\ \bibinfo {author} {\bibfnamefont {D.~J.}\ \bibnamefont
  {Thouless}},\ }\bibfield  {title} {\enquote {\bibinfo {title} {Ordering,
  metastability and phase transitions in two-dimensional systems},}\ }\href
  {\doibase 10.1088/0022-3719/6/7/010} {\bibfield  {journal} {\bibinfo
  {journal} {J. Phys. C: Solid State}\ }\textbf {\bibinfo {volume} {6}},\
  \bibinfo {pages} {1181} (\bibinfo {year} {1973})}\BibitemShut {NoStop}%
\bibitem [{\citenamefont {Benfatto}\ \emph {et~al.}(2008)\citenamefont
  {Benfatto}, \citenamefont {Capone}, \citenamefont {Caprara}, \citenamefont
  {Castellani},\ and\ \citenamefont {Di~Castro}}]{Benfatto2008}%
  \BibitemOpen
  \bibfield  {author} {\bibinfo {author} {\bibfnamefont {L.}~\bibnamefont
  {Benfatto}}, \bibinfo {author} {\bibfnamefont {M.}~\bibnamefont {Capone}},
  \bibinfo {author} {\bibfnamefont {S.}~\bibnamefont {Caprara}}, \bibinfo
  {author} {\bibfnamefont {C.}~\bibnamefont {Castellani}}, \ and\ \bibinfo
  {author} {\bibfnamefont {C.}~\bibnamefont {Di~Castro}},\ }\bibfield  {title}
  {\enquote {\bibinfo {title} {Multiple gaps and superfluid density from
  interband pairing in a four-band model of the iron oxypnictides},}\ }\href
  {\doibase 10.1103/PhysRevB.78.140502} {\bibfield  {journal} {\bibinfo
  {journal} {Phys. Rev. B}\ }\textbf {\bibinfo {volume} {78}},\ \bibinfo
  {pages} {140502} (\bibinfo {year} {2008})}\BibitemShut {NoStop}%
\bibitem [{\citenamefont {Botelho}\ and\ \citenamefont {S\'a~de
  Melo}(2006)}]{Botelho2006}%
  \BibitemOpen
  \bibfield  {author} {\bibinfo {author} {\bibfnamefont {S.~S.}\ \bibnamefont
  {Botelho}}\ and\ \bibinfo {author} {\bibfnamefont {C.~A.~R.}\ \bibnamefont
  {S\'a~de Melo}},\ }\bibfield  {title} {\enquote {\bibinfo {title}
  {Vortex-antivortex lattice in ultracold fermionic gases},}\ }\href {\doibase
  10.1103/PhysRevLett.96.040404} {\bibfield  {journal} {\bibinfo  {journal}
  {Phys. Rev. Lett.}\ }\textbf {\bibinfo {volume} {96}},\ \bibinfo {pages}
  {040404} (\bibinfo {year} {2006})}\BibitemShut {NoStop}%
\bibitem [{\citenamefont {Rist}\ \emph {et~al.}(2013)\citenamefont {Rist},
  \citenamefont {Varlamov}, \citenamefont {MacDonald}, \citenamefont {Fazio},\
  and\ \citenamefont {Polini}}]{Rist2013}%
  \BibitemOpen
  \bibfield  {author} {\bibinfo {author} {\bibfnamefont {S.}~\bibnamefont
  {Rist}}, \bibinfo {author} {\bibfnamefont {A.~A.}\ \bibnamefont {Varlamov}},
  \bibinfo {author} {\bibfnamefont {A.~H.}\ \bibnamefont {MacDonald}}, \bibinfo
  {author} {\bibfnamefont {R.}~\bibnamefont {Fazio}}, \ and\ \bibinfo {author}
  {\bibfnamefont {M.}~\bibnamefont {Polini}},\ }\bibfield  {title} {\enquote
  {\bibinfo {title} {Photoemission spectra of massless {D}irac fermions on the
  verge of exciton condensation},}\ }\href
  {https://journals.aps.org/prb/abstract/10.1103/PhysRevB.87.075418} {\bibfield
   {journal} {\bibinfo  {journal} {Phys. Rev. B}\ }\textbf {\bibinfo {volume}
  {87}},\ \bibinfo {pages} {075418} (\bibinfo {year} {2013})}\BibitemShut
  {NoStop}%
\bibitem [{\citenamefont {Miao}\ \emph {et~al.}(2012)\citenamefont {Miao},
  \citenamefont {Richard}, \citenamefont {Tanaka}, \citenamefont {Nakayama},
  \citenamefont {Qian}, \citenamefont {Umezawa}, \citenamefont {Sato},
  \citenamefont {Xu}, \citenamefont {Shi}, \citenamefont {Xu}, \citenamefont
  {Wang}, \citenamefont {Zhang}, \citenamefont {Yang}, \citenamefont {Xu},
  \citenamefont {Wen}, \citenamefont {Gu}, \citenamefont {Dai}, \citenamefont
  {Hu}, \citenamefont {Takahashi},\ and\ \citenamefont {Ding}}]{Miao2012}%
  \BibitemOpen
  \bibfield  {author} {\bibinfo {author} {\bibfnamefont {H.}~\bibnamefont
  {Miao}}, \bibinfo {author} {\bibfnamefont {P.}~\bibnamefont {Richard}},
  \bibinfo {author} {\bibfnamefont {Y.}~\bibnamefont {Tanaka}}, \bibinfo
  {author} {\bibfnamefont {K.}~\bibnamefont {Nakayama}}, \bibinfo {author}
  {\bibfnamefont {T.}~\bibnamefont {Qian}}, \bibinfo {author} {\bibfnamefont
  {K.}~\bibnamefont {Umezawa}}, \bibinfo {author} {\bibfnamefont
  {T.}~\bibnamefont {Sato}}, \bibinfo {author} {\bibfnamefont {Y.-M.}\
  \bibnamefont {Xu}}, \bibinfo {author} {\bibfnamefont {Y.~B.}\ \bibnamefont
  {Shi}}, \bibinfo {author} {\bibfnamefont {N.}~\bibnamefont {Xu}}, \bibinfo
  {author} {\bibfnamefont {X.-P.}\ \bibnamefont {Wang}}, \bibinfo {author}
  {\bibfnamefont {P.}~\bibnamefont {Zhang}}, \bibinfo {author} {\bibfnamefont
  {H.-B.}\ \bibnamefont {Yang}}, \bibinfo {author} {\bibfnamefont {Z.-J.}\
  \bibnamefont {Xu}}, \bibinfo {author} {\bibfnamefont {J.~S.}\ \bibnamefont
  {Wen}}, \bibinfo {author} {\bibfnamefont {G.-D.}\ \bibnamefont {Gu}},
  \bibinfo {author} {\bibfnamefont {X.}~\bibnamefont {Dai}}, \bibinfo {author}
  {\bibfnamefont {J.-P.}\ \bibnamefont {Hu}}, \bibinfo {author} {\bibfnamefont
  {T.}~\bibnamefont {Takahashi}}, \ and\ \bibinfo {author} {\bibfnamefont
  {H.}~\bibnamefont {Ding}},\ }\bibfield  {title} {\enquote {\bibinfo {title}
  {Isotropic superconducting gaps with enhanced pairing on electron fermi
  surfaces in {FeTe${}_{0.55}$Se${}_{0.45}$}},}\ }\href {\doibase
  10.1103/PhysRevB.85.094506} {\bibfield  {journal} {\bibinfo  {journal} {Phys.
  Rev. B}\ }\textbf {\bibinfo {volume} {85}},\ \bibinfo {pages} {094506}
  (\bibinfo {year} {2012})}\BibitemShut {NoStop}%
\bibitem [{\citenamefont {Daghero}\ \emph {et~al.}(2014)\citenamefont
  {Daghero}, \citenamefont {Pecchio}, \citenamefont {Ummarino}, \citenamefont
  {Nabeshima}, \citenamefont {Imai}, \citenamefont {Maeda}, \citenamefont
  {Tsukada}, \citenamefont {Komiya},\ and\ \citenamefont
  {Gonnelli}}]{Daghero2014}%
  \BibitemOpen
  \bibfield  {author} {\bibinfo {author} {\bibfnamefont {D.}~\bibnamefont
  {Daghero}}, \bibinfo {author} {\bibfnamefont {P.}~\bibnamefont {Pecchio}},
  \bibinfo {author} {\bibfnamefont {G.~A.}\ \bibnamefont {Ummarino}}, \bibinfo
  {author} {\bibfnamefont {F.}~\bibnamefont {Nabeshima}}, \bibinfo {author}
  {\bibfnamefont {Y.}~\bibnamefont {Imai}}, \bibinfo {author} {\bibfnamefont
  {A.}~\bibnamefont {Maeda}}, \bibinfo {author} {\bibfnamefont
  {I.}~\bibnamefont {Tsukada}}, \bibinfo {author} {\bibfnamefont
  {S.}~\bibnamefont {Komiya}}, \ and\ \bibinfo {author} {\bibfnamefont {R.~S.}\
  \bibnamefont {Gonnelli}},\ }\bibfield  {title} {\enquote {\bibinfo {title}
  {Point-contact {A}ndreev-reflection spectroscopy in {F}e ({T}e, {S}e) films:
  multiband superconductivity and electron-boson coupling},}\ }\href
  {https://iopscience.iop.org/article/10.1088/0953-2048/27/12/124014/meta}
  {\bibfield  {journal} {\bibinfo  {journal} {Supercond. Sci. Tech.}\ }\textbf
  {\bibinfo {volume} {27}},\ \bibinfo {pages} {124014} (\bibinfo {year}
  {2014})}\BibitemShut {NoStop}%
\bibitem [{\citenamefont {Kuzmicheva}\ \emph {et~al.}(2016)\citenamefont
  {Kuzmicheva}, \citenamefont {Kuzmichev}, \citenamefont {Sadakov},
  \citenamefont {Muratov}, \citenamefont {Usoltsev}, \citenamefont
  {Martovitsky}, \citenamefont {Shipilov}, \citenamefont {Chareev},
  \citenamefont {Mitrofanova},\ and\ \citenamefont {Pudalov}}]{Kuzmicheva2016}%
  \BibitemOpen
  \bibfield  {author} {\bibinfo {author} {\bibfnamefont {T.~E.}\ \bibnamefont
  {Kuzmicheva}}, \bibinfo {author} {\bibfnamefont {S.~A.}\ \bibnamefont
  {Kuzmichev}}, \bibinfo {author} {\bibfnamefont {A.~V.}\ \bibnamefont
  {Sadakov}}, \bibinfo {author} {\bibfnamefont {A.~V.}\ \bibnamefont
  {Muratov}}, \bibinfo {author} {\bibfnamefont {A.~S.}\ \bibnamefont
  {Usoltsev}}, \bibinfo {author} {\bibfnamefont {V.~P.}\ \bibnamefont
  {Martovitsky}}, \bibinfo {author} {\bibfnamefont {A.~R.}\ \bibnamefont
  {Shipilov}}, \bibinfo {author} {\bibfnamefont {D.~A.}\ \bibnamefont
  {Chareev}}, \bibinfo {author} {\bibfnamefont {E.~S.}\ \bibnamefont
  {Mitrofanova}}, \ and\ \bibinfo {author} {\bibfnamefont {V.~M.}\ \bibnamefont
  {Pudalov}},\ }\bibfield  {title} {\enquote {\bibinfo {title} {Direct evidence
  of two superconducting gaps in {FeSe$_{0.5}$ Te$_{0.5}$}: n{-A}ndreev
  spectroscopy and the lower critical field},}\ }\href
  {https://link.springer.com/article/10.1134/S0021364016240048} {\bibfield
  {journal} {\bibinfo  {journal} {JETP Lett.}\ }\textbf {\bibinfo {volume}
  {104}},\ \bibinfo {pages} {852} (\bibinfo {year} {2016})}\BibitemShut
  {NoStop}%
\bibitem [{\citenamefont {Yin}\ \emph {et~al.}(2015)\citenamefont {Yin},
  \citenamefont {Wu}, \citenamefont {Wang}, \citenamefont {Ye}, \citenamefont
  {Gong}, \citenamefont {Hou}, \citenamefont {Shan}, \citenamefont {Li},
  \citenamefont {Liang}, \citenamefont {Wu} \emph {et~al.}}]{Yin2015}%
  \BibitemOpen
  \bibfield  {author} {\bibinfo {author} {\bibfnamefont {J.~X.}\ \bibnamefont
  {Yin}}, \bibinfo {author} {\bibfnamefont {Z.}~\bibnamefont {Wu}}, \bibinfo
  {author} {\bibfnamefont {J.~H.}\ \bibnamefont {Wang}}, \bibinfo {author}
  {\bibfnamefont {Z.~Y.}\ \bibnamefont {Ye}}, \bibinfo {author} {\bibfnamefont
  {J.}~\bibnamefont {Gong}}, \bibinfo {author} {\bibfnamefont {X.~Y.}\
  \bibnamefont {Hou}}, \bibinfo {author} {\bibfnamefont {L.}~\bibnamefont
  {Shan}}, \bibinfo {author} {\bibfnamefont {A.}~\bibnamefont {Li}}, \bibinfo
  {author} {\bibfnamefont {X.~J.}\ \bibnamefont {Liang}}, \bibinfo {author}
  {\bibfnamefont {X.~X.}\ \bibnamefont {Wu}},  \emph {et~al.},\ }\bibfield
  {title} {\enquote {\bibinfo {title} {Observation of a robust zero-energy
  bound state in iron-based superconductor {F}e ({T}e, {S}e)},}\ }\href
  {https://www.nature.com/articles/nphys3371?draft=collection} {\bibfield
  {journal} {\bibinfo  {journal} {Nat. Phys.}\ }\textbf {\bibinfo {volume}
  {11}},\ \bibinfo {pages} {543} (\bibinfo {year} {2015})}\BibitemShut
  {NoStop}%
\bibitem [{\citenamefont {Perali}\ \emph {et~al.}(2002)\citenamefont {Perali},
  \citenamefont {Pieri}, \citenamefont {Strinati},\ and\ \citenamefont
  {Castellani}}]{Perali2002}%
  \BibitemOpen
  \bibfield  {author} {\bibinfo {author} {\bibfnamefont {A.}~\bibnamefont
  {Perali}}, \bibinfo {author} {\bibfnamefont {P.}~\bibnamefont {Pieri}},
  \bibinfo {author} {\bibfnamefont {G.~C.}\ \bibnamefont {Strinati}}, \ and\
  \bibinfo {author} {\bibfnamefont {C.}~\bibnamefont {Castellani}},\ }\bibfield
   {title} {\enquote {\bibinfo {title} {Pseudogap and spectral function from
  superconducting fluctuations to the bosonic limit},}\ }\href {\doibase
  10.1103/PhysRevB.66.024510} {\bibfield  {journal} {\bibinfo  {journal} {Phys.
  Rev. B}\ }\textbf {\bibinfo {volume} {66}},\ \bibinfo {pages} {024510}
  (\bibinfo {year} {2002})}\BibitemShut {NoStop}%
\bibitem [{\citenamefont {Salasnich}\ \emph {et~al.}(2019)\citenamefont
  {Salasnich}, \citenamefont {Shanenko}, \citenamefont {Vagov}, \citenamefont
  {Aguiar},\ and\ \citenamefont {Perali}}]{Salasnich2019}%
  \BibitemOpen
  \bibfield  {author} {\bibinfo {author} {\bibfnamefont {L.}~\bibnamefont
  {Salasnich}}, \bibinfo {author} {\bibfnamefont {A.~A.}\ \bibnamefont
  {Shanenko}}, \bibinfo {author} {\bibfnamefont {A.}~\bibnamefont {Vagov}},
  \bibinfo {author} {\bibfnamefont {J.~Albino}\ \bibnamefont {Aguiar}}, \ and\
  \bibinfo {author} {\bibfnamefont {A.}~\bibnamefont {Perali}},\ }\bibfield
  {title} {\enquote {\bibinfo {title} {Screening of pair fluctuations in
  superconductors with coupled shallow and deep bands: {A} route to
  higher-temperature superconductivity},}\ }\href {\doibase
  10.1103/PhysRevB.100.064510} {\bibfield  {journal} {\bibinfo  {journal}
  {Phys. Rev. B}\ }\textbf {\bibinfo {volume} {100}},\ \bibinfo {pages}
  {064510} (\bibinfo {year} {2019})}\BibitemShut {NoStop}%
\end{thebibliography}
%

\end{document}